\documentclass[12pt,aps,pra,showkeys]{revtex4}
\usepackage{amsmath}
\usepackage{amscd}
\usepackage{graphicx}
\usepackage{cases}

\begin{document}

\title[Short Title]{Complete Bell-state analysis for superconducting-quantum-interference-device qubits with transitionless tracking algorithm}

\author{Yi-Hao Kang$^{1,2}$}
\author{Ye-Hong Chen$^{1,2}$}
\author{Zhi-Cheng Shi$^{1,2}$}
\author{Bi-Hua Huang$^{1,2}$}
\author{Jie Song$^{3}$}
\author{Yan Xia$^{1,2,}$\footnote{E-mail: xia-208@163.com}}

\affiliation{$^{1}$Department of Physics, Fuzhou University, Fuzhou 350116, China\\
             $^{2}$Fujian Key Laboratory of Quantum Information and Quantum Optics (Fuzhou University), Fuzhou 350116, China\\
             $^{3}$Department of Physics, Harbin Institute of Technology, Harbin 150001, China}

\begin{abstract}

In this paper, we propose a protocol for complete Bell-state
analysis for two superconducting-quantum-interference-device qubits.
The Bell-state analysis could be completed by using a sequence of
microwave pulses designed by the transitionless tracking algorithm,
which is an useful method in the technique of shortcut to
adiabaticity. After the whole process, the information for
distinguishing four Bell states will be encoded on two auxiliary
qubits, while the Bell states keep unchanged. One can read out the
information by detecting the auxiliary qubits. Thus the Bell-state
analysis is nondestructive. The numerical simulations show that the
protocol possesses high success probability of distinguishing each
Bell state with current experimental technology even when
decoherence is taken into account. Thus, the protocol may have
potential applications for the information readout in quantum
communications and quantum computations in superconducting quantum
networks.

\end{abstract}

\maketitle

\section{Introduction}

Entanglement is a basic concept in quantum information science. It
provides possibility to test quantum nonlocality against local
hidden theory \cite{BellPhysics1,Greenberger,DurPRA62}, and also
plays a key role in various quantum information tasks
\cite{KarlssonPRA58,DFGPRA72,EkertPRL67,DFGPRA68,BennettPRL69,LXSPRA65,SYBPRA81I}.
Therefore, preparing \cite{DZJPRA74,DLMPRL90}, transferring
\cite{WTJPRA85,HCYPRB83} and purifying \cite{RBCPRA90,PanNat410} all
kinds of entangled states in different physical systems become hot
topics in quantum information processing (QIP). As Bell states of
two qubits are easy to be obtained and manipulated, they have been
employed as the information carriers in quantum communications and
quantum computations \cite{EkertPRL67,BennettPRL69,BennettPRL68}.
Thus when using Bell states as information carriers, reading out
quantum information encoded on Bell states is an indispensable task,
which greatly motivated the researches on the Bell-state analysis.
At the beginning, researchers mainly paid their attentions on the
Bell-state analysis for polarized photons with liner optical
elements \cite{MattlePRL76,HouwelingenPRL96}. But unfortunately, it
have been proven by protocols \cite{VaidmanPRA59,CalsamigliaPRA65}
that the Bell-state analysis with only linear optical element have
optimal success probability of 0.5. Besides, the Bell-state analysis
usually destroys the entanglement which causes the waste of physical
resources. Therefore, to achieve complete and nondestructive
Bell-state analysis and to exploit the advantages of other physical
systems, researchers have turned their attentions on Bell states in
various systems by applying many new techniques, such as
nonlinearities and hyperentanglement. Until now, complete and
nondestructive Bell-state analysis for photons
\cite{SYBPRA81II,SYBPRA82,BarbieriPRA75,WTJPRA86,RBCOE20,BonatoPRL104,XYJOSAB31},
atoms \cite{HYCPB19}, spins inside quantum dots
\cite{WHRIJTP52,KYHAPB119} and nitrogen-vacancy centers
\cite{LJZIJTP56} have been reported.

In recent years, the superconducting system has been developed a
lot, and is now deemed as a very promising candidate to implement
quantum information tasks
\cite{MakhlinRMP73,XZLRMP85,VionSci296,YYSci296,YCPPRL92,YCPPRA67,YCPPRA74,YCPPRA82,ClarkeNature453,DevoretADP16,YCPPRA86,BlaisPRA69,WallraffNature431,YCPPRA87,ChiorescuNature431,SteinbachPRL87,FilippPRL102,BialczakNP6,YamamotoPRB82,ReedPRL105,MajerNature449,DiCarloNature460,SchmidtADP525,StrauchPRL105,KochPRA76},
as it possesses many advantages. Superconducting qubits, including
phase qubits, change qubits, flux qubits, etc., are outstanding with
their relatively long decoherence time \cite{ClarkeNature453} and
perfect scalability \cite{VionSci296,YYSci296,ChiorescuNature431}.
Among all kinds of superconducting qubits, the
superconducting-quantum-interference-device (SQUID) qubits in cavity
quantum electrodynamics (QED) have many advantages:

(1) The positions of SQUID qubits in a cavity are fixed. That makes
them holds superiority compared with neutral atoms, which requires
to be controlled the centers of mass motion in a cavity.
\cite{YCPPRL92,YCPPRA67}.

(2) When placing SQUID qubits into a superconducting cavity,
decoherence induced due to the external environment can be greatly
suppressed since the superconducting cavity could be considered as
the magnetic shield for SQUID qubits \cite{YCPPRA67}.

(3) The strong-coupling limit of the cavity QED can be easily
realized for SQUID qubits embedded in a cavity, while it is
difficult to be realized with atoms \cite{YCPPRL92}.

(4) The level structure of every individual SQUID qubit can be
adjusted easily \cite{YCPPRL92}.

The great advantages of SQUID qubits make them attractive choices to
implement quantum information tasks. So far, SQUID qubits have been
widely used in entanglement preparations
\cite{YCPPRL92,YCPPRA67,DZJPRA74,SKHPRA75}, information transfers
\cite{YCPPRL92,YCPPRA67}, logic gates \cite{YCPPRA67}. However,
Bell-state analysis for SQUID qubits still has plenty room for
researches.

On the other hand, when choosing superconducting system as the
platform for QIP, an ineluctable question is to design microwave
pulses driving superconducting qubits to complete various
operations. Interestingly, a new technique called by shortcut to
adiabaticity (STA)
\cite{TorronteguiAAMOP62,BerryJPA42,CXPRL105,CYHSR6,CampoPRL111,CampoPRL109,CXPRA83,MugaJPB42,CXPRL104,CampoSR2,BaksicPRL116,GaraotPRA89,SaberiPRA90,TorronteguiPRA89,TorosovPRA87,TorosovPRA89,IbanezPRA87,IbanezPRL109,SXKPRA93,TorronteguiPRA83,MugaJPB43,TorronteguiPRA85,MasudaPRA84,MasudaJPCA119,CXPRA82,CXPRA84,CampoPRA84,CampoEPL96,SchaffPRA82,CXPRA86,SantosPRA93,HenPRA91,DeffnerPRX4,SXKNJP6}
has been developing recently to control quantum evolutions. Rather
than confining quantum evolutions along one eigenstate or
superpositions of several eigenstates of the Hamiltonian under the
adiabatic condition, STA provides a lot of evolution paths by means
of various methods including transitionless tracking algorithm
\cite{BerryJPA42,CXPRL105,CXPRA83,CampoPRL111,CYHSR6},
Lewis-Riesenfeld invariants theory \cite{CXPRA83,CXPRL104}, Lie
algebra \cite{GaraotPRA89,TorronteguiPRA89}, picture transformations
\cite{BaksicPRL116,IbanezPRA87,IbanezPRL109,SXKPRA93}, fast-forward
scales \cite{MasudaPRA84,MasudaJPCA119}, etc.. These protocols
\cite{TorronteguiAAMOP62,BerryJPA42,CXPRL105,CYHSR6,CampoPRL111,CampoPRL109,CXPRA83,MugaJPB42,CXPRL104,CampoSR2,BaksicPRL116,GaraotPRA89,SaberiPRA90,TorronteguiPRA89,TorosovPRA87,TorosovPRA89,IbanezPRA87,IbanezPRL109,SXKPRA93,TorronteguiPRA83,MugaJPB43,TorronteguiPRA85,MasudaPRA84,MasudaJPCA119,CXPRA82,CXPRA84,CampoPRA84,CampoEPL96,SchaffPRA82,CXPRA86,SantosPRA93,HenPRA91,DeffnerPRX4,SXKNJP6}
have demonstrated that STA not only inherits the robustness of the
adiabatic passage, but also greatly accelerates adiabatic processes.
Moreover, constructing STA by using different methods produces
excellent feasibility to handle all kinds of quantum information
tasks. Thus, it may be a good idea applying STA in pulse design to
manipulate superconducting systems.

In this paper, motivated by (1) the importance of Bell-state
analysis in quantum information tasks, (2) the advantages of SQUID
qubits, (3) the requirement of Bell-state analysis from quantum
communications and computations within superconducting quantum
networks, (4) the advantages of STA in designing pulses to control
physical systems, we proposed a protocol for complete and
nondestructive Bell state analysis for two SQUID qubits. By using
transitionless tracking algorithm, a useful method of STA, a
sequence of microwave pulses are designed to complete the Bell-state
analysis. The information for distinguishing four Bell states would
be encoded in two auxiliary SQUID qubits, and could be read out with
current technology \cite{BennettSST20,TakayanagiSM32}. Therefore,
the operations of the Bell-state analysis are not difficult in real
experiments. Besides, the protocol combines the robustness of SQUID
qubits and the speediness of STA. Thus, we can see in numerical
simulation that high success probability to distinguish each Bell
state are still available when decoherence is considered. By
substituting experimentally realizable parameters, good performance
of the Bell-state analysis is shown.

The article is organized as follows. In Sec. II, we briefly review
the physical model of a SQUID qubit. In Sec. III, we amply
illuminate the procedures of the Bell-state analysis. In Sec. IV,
the transitionless tracking algorithm is utilized to design a
microwave pulses for realizing the Bell-state analysis. In Sec. V,
numerical simulations are performed to select suitable control
parameters and demonstrate the robustness of the Bell-state analysis
against decoherence. Finally, conclusions are given in Sec. VI.

\section{Physical model of a SQUID qubit}

\begin{figure}
\scalebox{0.6}{\includegraphics[scale=1]{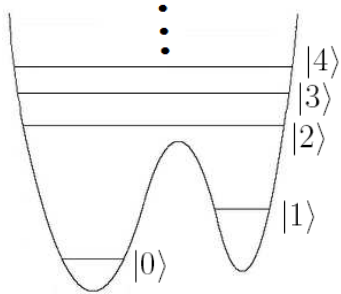}} \caption{The
level configuration of a single SQUID qubit.}\label{fig1}
\end{figure}

Considering a single SQUID qubit with junction capacitance $C$ and
loop inductance $L$, its Hamiltonian reads \cite{YCPPRA67,YCPPRL92}
\begin{eqnarray}\label{e1}
H_{s}(t)=\frac{Q^2}{2C}+\frac{(\Phi-\Phi_x)^2}{2L}-E_J\cos(2\pi\frac{\Phi}{\Phi_0}),
\end{eqnarray}
where, $Q$ is the total charge on the capacitor; $\Phi$ is the
magnetic flux threading the loop, and $\Phi_x$ is the external flux
applied to the ring; $E_J=I_c\Phi_0/2\pi$ is the Josephson energy
with $I_c$ and $\Phi_0=h/2e$ being the critical current of the
junction and the flux quantum. By quantizing the Hamiltonian of the
SQUID qubit, the SQUID qubit can be described by level diagram with
a serial of energy levels $\{|k\rangle\}$ ($k=0,1,2,...$) shown in
Fig.~\ref{fig1} \cite{YCPPRA67,YCPPRL92}.
When a transition between two different levels $|k\rangle$ and
$|k'\rangle$ is driving by an classical microwave field, in the
frame of the rotating-wave approximation, the Rabi frequency of the
driving field could be written by \cite{YCPPRA67,YCPPRL92}
\begin{eqnarray}\label{e2}
\Omega_{kk'}(t)=\frac{1}{2L\hbar}\langle
k|\Phi|k'\rangle\int_{S}\tilde{\mathbf{B}}_{\mu
w}(\mathbf{r},t)\cdot d\mathbf{S},
\end{eqnarray}
where, $S$ is surface bounded by the loop of the SQUID qubit;
$\mathbf{B}_{\mu w}(\mathbf{r},t)=\tilde{\mathbf{B}}_{\mu
w}(\mathbf{r},t)\cos(2\pi\nu_{\mu w}t)$ is the magnetic components
of the classical microwave in the superconducting loop of the SQUID
qubit with frequency $\nu_{\mu w}$. Considering that the SQUID qubit
is placed in a microwave cavity, when a transition between two
different levels $|k\rangle$ and $|k'\rangle$ is coupled to a
quantized cavity field with frequency $\omega_c$, after the
rotating-wave approximation, the coupling constant reads
\cite{YCPPRA67,YCPPRL92}
\begin{eqnarray}\label{e3}
g_{kk'}=\frac{1}{L}\sqrt{\frac{\omega_c}{2\mu_0\hbar}}\langle
k|\Phi|k'\rangle\int_{S}\mathbf{B}_c(\mathbf{r})\cdot d\mathbf{S},
\end{eqnarray}
where, $\mathbf{B}_c(\mathbf{r})$ is the magnetic components of the
cavity mode in the superconducting loop of the SQUID qubit.

\section{Complete Bell-state analysis}

\begin{figure}
\scalebox{0.6}{\includegraphics[scale=0.9]{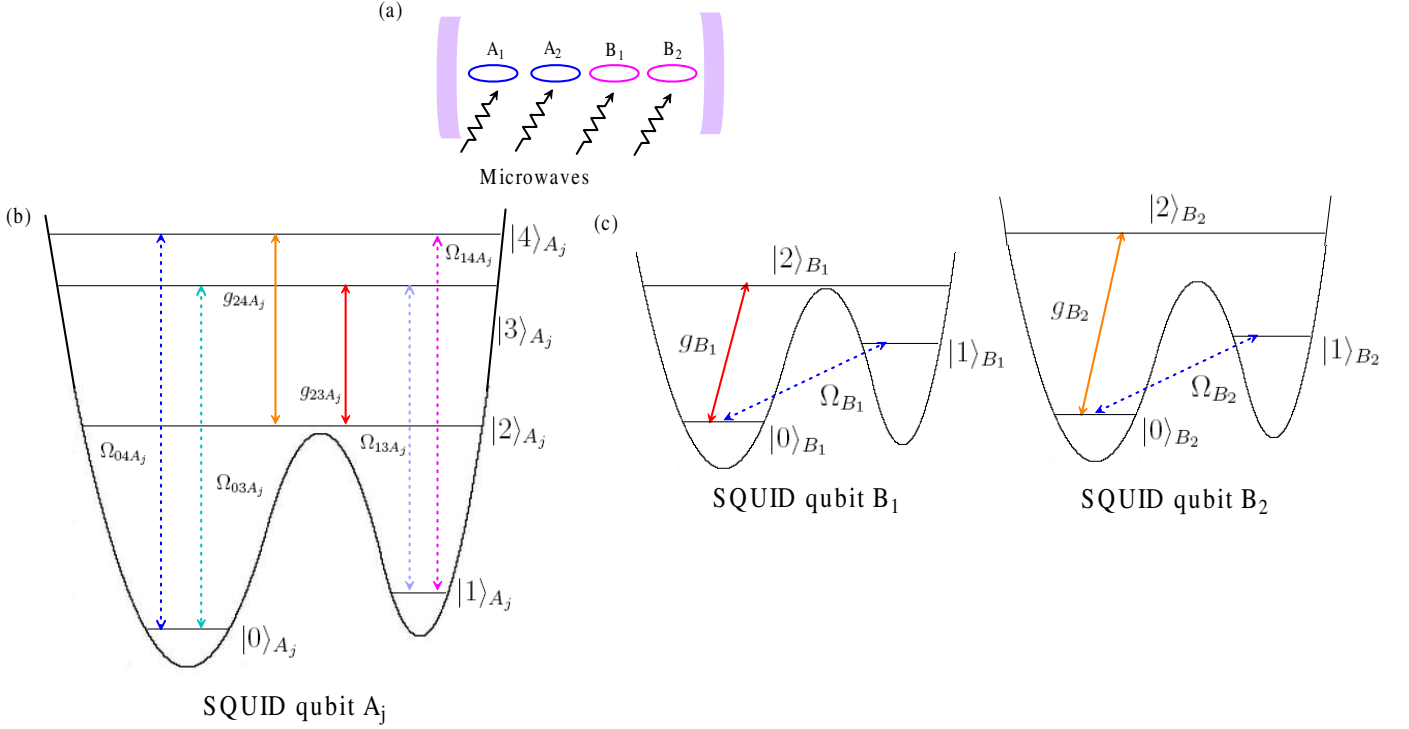}} \caption{(a)
The auxiliary qubits $A_1$, $A_2$ and the information carriers $B_1$
and $B_2$ placed in a microwave cavity. (b) The level configuration
of SQUID qubit $A_j$ ($j=1,2$). (c) The level configurations of
SQUID qubits $B_1$ and $B_2$.}\label{fig2}
\end{figure}

Consider a system that contains four SQUID qubits $A_1$, $A_2$,
$B_1$, $B_2$ placing inside a microwave cavity, which is shown in
Fig.~\ref{fig2}~(a). SQUID qubit $A_j$ ($j=1,2$) is employed as an
auxiliary qubit, whose level diagram is shown in
Fig.~\ref{fig2}~(b). We consider the lowest five levels
$|0\rangle_{A_j}$, $|1\rangle_{A_j}$, $|2\rangle_{A_j}$,
$|3\rangle_{A_j}$, and $|4\rangle_{A_j}$ of SQUID qubit $A_j$. The
transition between $|0\rangle_{A_j}$ and $|3\rangle_{A_j}$
($|4\rangle_{A_j}$) is resonantly driven by a classical microwave
field with Rabi frequency $\Omega_{03A_j}(t)$ ($\Omega_{04A_j}(t)$).
The transition between $|1\rangle_{A_j}$ and $|3\rangle_{A_j}$
($|4\rangle_{A_j}$) is resonantly driven by a classical microwave
field with Rabi frequency $\Omega_{13A_j}(t)$ ($\Omega_{14A_j}(t)$).
SQUID qubits $B_1$ and $B_2$ are information carriers, whose level
diagrams are shown in Fig.~\ref{fig2}~(c). We consider the lowest
three levels $|0\rangle_{B_j}$, $|1\rangle_{B_j}$, $|2\rangle_{B_j}$
and among them, information is encoded on $|0\rangle_{B_j}$ and
$|1\rangle_{B_j}$. Thus, the four Bell states to be distinguished
can be described as
\begin{eqnarray}\label{e4}
&&|\Psi_{\pm}\rangle_{B_1B_2}=\frac{1}{2}(|0\rangle_{B_1}|0\rangle_{B_2}\pm|1\rangle_{B_1}|1\rangle_{B_2}),\cr\cr&&
|\Phi_{\pm}\rangle_{B_1B_2}=\frac{1}{2}(|0\rangle_{B_1}|1\rangle_{B_2}\pm|1\rangle_{B_1}|0\rangle_{B_2}).
\end{eqnarray}
A classical microwave field with Rabi frequency $\Omega_{B_j}(t)$ is
applied on SQUID qubit $B_j$ to drive the transition between levels
$|0\rangle_{B_j}$ and $|1\rangle_{B_j}$.
Assuming the microwave cavity is a double-mode cavity, where two
cavity fields $a_1$ and $a_2$ (denote by their annihilation
operators) may exist. The cavity field $a_1$ ($a_2$) could
resonantly coupled with the transition between levels
$|2\rangle_{A_j}$ ($j=1,2$) and $|3\rangle_{A_j}$ ($|2\rangle_{A_j}$
and $|4\rangle_{A_j}$) of SQUID qubit $A_j$ with coupling constant
$g_{23A_j}$ ($g_{24A_j}$), and the transition between levels
$|0\rangle_{B_1}$ and $|2\rangle_{B_1}$ ($|0\rangle_{B_2}$ and
$|2\rangle_{B_2}$) with coupling constant $g_{B_1}$ ($g_{B_2}$).
Assuming the frequencies of cavity modes $a_1$ and $a_2$ are
$\omega_1$ and $\omega_2$, respectively. The transition frequency
between $|0\rangle_{B_1}$ and $|2\rangle_{B_1}$ ($|0\rangle_{B_2}$
and $|2\rangle_{B_2}$) of SQUID qubit $B_1$ ($B_2$) should be equal
to $\omega_1$ ($\omega_2$). According to Ref.~\cite{YCPPRL92}, level
structure of each individual SQUID qubit can be adjusted by either
design variations and/or changing local bias field. Thus, coupling
between microwave pulses (cavity fields) and any particular SQUID
qubits can be obtained selectively via frequency matching. When
$g_{B_1},g_{B_2}\ll|\omega_1-\omega_2|$, the interaction between
SQUID qubit $B_1$ and cavity mode $a_2$ (SQUID qubit $B_2$ and
cavity mode $a_1$) could be discarded \cite{YCPPRA67}. Therefore, in
the interaction picture, the total Hamiltonian of system for
Bell-state analysis could be written as
\begin{eqnarray}\label{e5}
&&H_{I}(t)=H_{mA}(t)+H_{mB}+H_{c},\cr\cr&&
H_{mA}(t)=\sum\limits_{j=1,2}\Omega_{03A_j}(t)|0\rangle\langle3|+\Omega_{04A_j}(t)|0\rangle\langle4|
+\Omega_{13A_j}(t)|1\rangle\langle3|+\Omega_{14A_j}(t)|1\rangle\langle4|+H.c.,\cr\cr&&
H_{mB}(t)=\sum\limits_{j=1,2}\Omega_{B_j}(t)e^{-i\epsilon_j}|0\rangle\langle1|+H.c.,\cr\cr&&
H_{c}=\sum\limits_{j=1,2}g_{23A_j}|3\rangle_{A_j}\langle2|a_1+g_{24A_j}|4\rangle_{A_j}\langle2|a_2+g_{B_j}|2\rangle_{B_j}\langle0|a_j+H.c.,
\end{eqnarray}
where, $\epsilon_j$ is the phase shift of $\Omega_{B_j}$. Here, we
take $\epsilon_j=\pi/2$ for the convenience of calculations and
descriptions.

Now, let us describe the procedures for Bell-state analysis. The
Bell-state can be divided into six steps. We do not discuss the
pulse design here, but leave it later in Sec. IV. Besides, for the
convenience of descriptions, we assume the operation time of each
step is $T$.

\textbf{Step 1:} Assume SQUID qubit $A_j$ is initially in state
$|0\rangle_{A_j}$, cavity field $a_j$ is initially in vacuum state
$|0\rangle_{a_j}$. We turn on $\Omega_{03A_1}(t)$ and
$\Omega_{13A_1}(t)$, but turn off other classical microwave fields.
In this case, SQUID qubit $A_2$ is decoupled to the system. Besides,
cavity fields $a_2$ keeps in vacuum state. Thus, whether SQUID qubit
$B_2$ is in state $|0\rangle_{B_2}$ or $|1\rangle_{B_2}$, it does
not evolve as well. Without considering cavity field $a_2$ and the
decoupled SQUID qubits, the system would evolve in a subspace
spanned by
\begin{eqnarray}\label{e6}
&&|\bar{\psi}_1\rangle=|0\rangle_{A_1}|0\rangle_{B_1}|0\rangle_{a_1},\
\
|\bar{\psi}_2\rangle=|3\rangle_{A_1}|0\rangle_{B_1}|0\rangle_{a_1},\
\
|\bar{\psi}_3\rangle=|2\rangle_{A_1}|0\rangle_{B_1}|1\rangle_{a_1},\cr\cr
&&|\bar{\psi}_4\rangle=|2\rangle_{A_1}|2\rangle_{B_1}|0\rangle_{a_1},\
\
|\bar{\psi}_5\rangle=|1\rangle_{A_1}|0\rangle_{B_1}|0\rangle_{a_1},\
\
|\bar{\psi}_6\rangle=|0\rangle_{A_1}|1\rangle_{B_1}|0\rangle_{a_1},\cr\cr
&&|\bar{\psi}_7\rangle=|3\rangle_{A_1}|1\rangle_{B_1}|0\rangle_{a_1},\
\
|\bar{\psi}_8\rangle=|2\rangle_{A_1}|1\rangle_{B_1}|1\rangle_{a_1},\
\
|\bar{\psi}_9\rangle=|1\rangle_{A_1}|1\rangle_{B_1}|0\rangle_{a_1},
\end{eqnarray}
where $|1\rangle_{a_1}$ denotes the one-photon state of cavity field
$a_1$. Rewriting the Hamiltonian of the system within the subspace,
we obtain
\begin{eqnarray}\label{e7}
&&H_{step1}(t)=H_{m1}(t)+H_{c1},\cr\cr&&
H_{m1}(t)=\Omega_{03A_1}(t)(|\bar{\psi}_1\rangle\langle\bar{\psi}_2|+|\bar{\psi}_6\rangle\langle\bar{\psi}_7|)+\Omega_{13A_1}(t)(|\bar{\psi}_5\rangle\langle\bar{\psi}_2|+|\bar{\psi}_9\rangle\langle\bar{\psi}_7|)+H.c.,\cr\cr&&
H_{c1}=g_{23A_1}(|\bar{\psi}_2\rangle\langle\bar{\psi}_3|+|\bar{\psi}_7\rangle\langle\bar{\psi}_8|)+g_{B_1}|\bar{\psi}_4\rangle_{B_1}\langle\bar{\psi}_3|+H.c..
\end{eqnarray}
The eigenstates of $H_{c1}$ are
\begin{eqnarray}\label{e8}
&&|\bar{\phi}_{0}\rangle=\frac{1}{\sqrt{g_{23A_1}^2+g_{B_1}^2}}(g_{B_1}|\bar{\psi}_2\rangle-g_{23A_1}|\bar{\psi}_4\rangle),\cr\cr
&&|\bar{\phi}_{1}\rangle=\frac{1}{\sqrt{2}}(|\bar{\psi}_7\rangle+|\bar{\psi}_8\rangle),\cr\cr
&&|\bar{\phi}_{2}\rangle=\frac{1}{\sqrt{2}}(|\bar{\psi}_7\rangle-|\bar{\psi}_8\rangle),\cr\cr
&&|\bar{\phi}_{3}\rangle=\frac{1}{\sqrt{2(g_{23A_1}^2+g_{B_1}^2)}}(g_{23A_1}|\bar{\psi}_2\rangle+\sqrt{g_{23A_1}^2+g_{B_1}^2}|\bar{\psi}_3\rangle+g_{B_1}|\bar{\psi}_4\rangle),\cr\cr
&&|\bar{\phi}_{4}\rangle=\frac{1}{\sqrt{2(g_{23A_1}^2+g_{B_1}^2)}}(g_{23A_1}|\bar{\psi}_2\rangle-\sqrt{g_{23A_1}^2+g_{B_1}^2}|\bar{\psi}_3\rangle+g_{B_1}|\bar{\psi}_4\rangle),
\end{eqnarray}
with corresponding eigenvalues 0, $g_{23A_1}$, $-g_{23A_1}$,
$\sqrt{g_{23A_1}^2+g_{B_1}^2}$, $-\sqrt{g_{23A_1}^2+g_{B_1}^2}$,
respectively. With the condition
$\Omega_{03A_1}(t),\Omega_{13A_1}(t)\ll g_{23A_1},g_{B_1}$, we can
derive the effective Hamiltonian of the system as
\begin{eqnarray}\label{e9}
H_{eff1}(t)=\frac{g_{B_1}}{\sqrt{g_{23A_1}^2+g_{B_1}^2}}[\Omega_{03A_1}(t)|\bar{\psi}_1\rangle\langle\bar{\phi}_0|+\Omega_{13A_1}(t)|\bar{\psi}_5\rangle\langle\bar{\phi}_0|]+H.c..
\end{eqnarray}
The details of the derivation of $H_{eff1}(t)$ are given in the
appendix. To realize
$|\bar{\psi}_1\rangle\rightarrow|\bar{\psi}_5\rangle$, we should
suitably design $\Omega_{03A_1}(t)$ and $\Omega_{13A_1}(t)$. The
design of $\Omega_{03A_1}(t)$ and $\Omega_{13A_1}(t)$ is amply
discussed in Sec. IV. Thus, if SQUID qubit $B_1$ is initially in
state $|0\rangle_{B_1}$, after Step 1, its state keeps unchanged,
while SQUID qubit $A_1$ evolves from $|0\rangle_{A_1}$ to
$|1\rangle_{A_1}$. But if qubit $B_1$ is initially in state
$|1\rangle_{B_1}$, both SQUID qubits $A_1$ and $B_1$ keep in their
initial states after Step 1.

\textbf{Step 2:} In this step, we turn on $\Omega_{04A_1}(t)$ and
$\Omega_{14A_1}(t)$, but turn off other classical microwave fields.
Similar to Step 1, SQUID qubit $A_2$ is decoupled to the system,
while cavity fields $a_1$ keeps in vacuum state. Thus, SQUID qubit
$B_1$ is decoupled to the system in this step. The system would
evolve in a subspace spanned by
\begin{eqnarray}\label{e10}
&&|\tilde{\psi}_1\rangle=|0\rangle_{A_1}|0\rangle_{B_2}|0\rangle_{a_2},\
\
|\tilde{\psi}_2\rangle=|4\rangle_{A_1}|0\rangle_{B_2}|0\rangle_{a_2},\
\
|\tilde{\psi}_3\rangle=|2\rangle_{A_1}|0\rangle_{B_2}|1\rangle_{a_2},\cr\cr
&&|\tilde{\psi}_4\rangle=|2\rangle_{A_1}|2\rangle_{B_2}|0\rangle_{a_2},\
\
|\tilde{\psi}_5\rangle=|1\rangle_{A_1}|0\rangle_{B_2}|0\rangle_{a_2},\
\
|\tilde{\psi}_6\rangle=|0\rangle_{A_1}|1\rangle_{B_2}|0\rangle_{a_2},\cr\cr
&&|\tilde{\psi}_7\rangle=|4\rangle_{A_1}|1\rangle_{B_2}|0\rangle_{a_2},\
\
|\tilde{\psi}_8\rangle=|2\rangle_{A_1}|1\rangle_{B_2}|1\rangle_{a_2},\
\
|\tilde{\psi}_9\rangle=|1\rangle_{A_1}|1\rangle_{B_2}|0\rangle_{a_2}.
\end{eqnarray}
Similar way, under the condition
$\Omega_{04A_1}(t),\Omega_{14A_1}(t)\ll g_{24A_1},g_{B_2}$, the
effective Hamiltonian of the system can be derived as
\begin{eqnarray}\label{e11}
H_{eff2}(t)=\frac{g_{B_2}}{\sqrt{g_{24A_1}^2+g_{B_2}^2}}[\Omega_{04A_1}(t)|\tilde{\psi}_1\rangle\langle\tilde{\phi}_0|+\Omega_{14A_1}(t)|\tilde{\psi}_5\rangle\langle\tilde{\phi}_0|]+H.c.,
\end{eqnarray}
with
$|\tilde{\phi}_{0}\rangle=\frac{1}{\sqrt{g_{24A_1}^2+g_{B_2}^2}}(g_{B_2}|\tilde{\psi}_2\rangle-g_{24A_1}|\tilde{\psi}_4\rangle)$.
The design of $\Omega_{04A_1}(t)$ and $\Omega_{14A_1}(t)$ for
achieving
$|\tilde{\psi}_1\rangle\leftrightarrow|\tilde{\psi}_5\rangle$ is
also discussed in Sec. IV. Therefore, if SQUID qubit $B_2$ is
initially in state $|0\rangle_{B_2}$, its state keeps unchanged,
while the evolution of SQUID qubit $A_1$ would be
$|0\rangle_{A_1}\rightarrow|1\rangle_{A_1}$ or
$|1\rangle_{A_1}\rightarrow|0\rangle_{A_1}$, where the initial state
of SQUID qubit $A_1$ in this step is decided by the result of Step
1. Otherwise, the initial state of SQUID qubit $B_2$ is
$|1\rangle_{B_2}$, both SQUID qubits $A_1$ and $B_2$ stay in their
initial states.

Step 1 and Step 2 can be summarized as Table I.
\begin{center}
\centering{\bf Table I. The evolution of $A_1$ in Step 1 and Step
2.} {\small\begin{tabular}{ccc} \hline\hline
\ \ \ \ The state of $B_1$ and $B_2$\ \ \ \ &\ \ \ \ Step 1\ \ \ \ &\ \ \ \ Step 2\ \ \ \ \\
\hline
\ \ \ \ $|0\rangle_{B_1}|0\rangle_{B_2}$\ \ \ \ &\ \ \ \ $|0\rangle_{A_1}\rightarrow|1\rangle_{A_1}$\ \ \ \ &\ \ \ \ $|1\rangle_{A_1}\rightarrow|0\rangle_{A_1}$\ \ \ \ \\
\ \ \ \ $|0\rangle_{B_1}|1\rangle_{B_2}$\ \ \ \ &\ \ \ \ $|0\rangle_{A_1}\rightarrow|1\rangle_{A_1}$\ \ \ \ &\ \ \ \ $|1\rangle_{A_1}\rightarrow|1\rangle_{A_1}$\ \ \ \ \\
\ \ \ \ $|1\rangle_{B_1}|0\rangle_{B_2}$\ \ \ \ &\ \ \ \ $|0\rangle_{A_1}\rightarrow|0\rangle_{A_1}$\ \ \ \ &\ \ \ \ $|0\rangle_{A_1}\rightarrow|1\rangle_{A_1}$\ \ \ \ \\
\ \ \ \ $|1\rangle_{B_1}|1\rangle_{B_2}$\ \ \ \ &\ \ \ \ $|0\rangle_{A_1}\rightarrow|0\rangle_{A_1}$\ \ \ \ &\ \ \ \ $|0\rangle_{A_1}\rightarrow|0\rangle_{A_1}$\ \ \ \ \\
\hline \hline
\end{tabular}}
\end{center}
Thus, after Step 1 and Step 2, information for distinguishing
$|\Psi_\pm\rangle_{B_1B_2}$ from $|\Phi_\pm\rangle_{B_1B_2}$ is
encoded on the SQUID qubit $A_1$.

\textbf{Step 3:} In this step, we only turn on classical microwave
fields $\Omega_{B_1}(t)$ and $\Omega_{B_2}(t)$ to perform
single-qubit operation on SQUID qubits $B_1$ and $B_2$. When
\begin{eqnarray}\label{e12}
\int_{2T}^{3T}\Omega_{B_j}(t')dt'=\pi/4,
\end{eqnarray}
the single-qubit operation on SQUID qubit $B_j$ is
\begin{eqnarray}\label{e13}
|0\rangle_{B_j}\rightarrow\frac{1}{\sqrt{2}}(|0\rangle_{B_j}+|1\rangle_{B_j}),\
\ \
|1\rangle_{B_j}\rightarrow-\frac{1}{\sqrt{2}}(|0\rangle_{B_j}-|1\rangle_{B_j}).
\end{eqnarray}
Therefore, the four Bell states change as follows
\begin{eqnarray}\label{e14}
&&|\Psi_+\rangle_{B_1B_2}\rightarrow|\Psi_+\rangle_{B_1B_2},\ \ \
|\Psi_-\rangle_{B_1B_2}\rightarrow|\Phi_+\rangle_{B_1B_2},\cr\cr
&&|\Phi_+\rangle_{B_1B_2}\rightarrow-|\Psi_-\rangle_{B_1B_2},\ \
|\Phi_-\rangle_{B_1B_2}\rightarrow|\Phi_-\rangle_{B_1B_2}.
\end{eqnarray}
Thus, we can transform the question of distinguishing
$|\Psi_+\rangle_{B_1B_2}$ ($|\Phi_+\rangle_{B_1B_2}$) from
$|\Psi_-\rangle_{B_1B_2}$ ($|\Phi_-\rangle_{B_1B_2}$) to the
question of distinguishing $|\Psi_\pm\rangle_{B_1B_2}$ from
$|\Phi_\pm\rangle_{B_1B_2}$.

\textbf{Step 4:} We turn on $\Omega_{03A_2}(t)$ and
$\Omega_{13A_2}(t)$, but turn off other classical microwave fields.
In this step, SQUID qubit $A_2$ takes the place of SQUID qubit $A_1$
in the Step 1. Thus, by suitably designing $\Omega_{03A_2}(t)$ and
$\Omega_{13A_2}(t)$, we can make SQUID qubit $A_2$ evolve from
$|0\rangle_{A_2}$ to $|1\rangle_{A_2}$ when SQUID qubit $B_1$ is in
state $|0\rangle_{B_1}$, while keep in $|0\rangle_{A_2}$ when qubit
$B_1$ is in state $|1\rangle_{B_1}$.

\textbf{Step 5:} We only turn on $\Omega_{04A_2}(t)$ and
$\Omega_{14A_2}(t)$. In this step, SQUID qubit $A_2$ takes the place
of SQUID qubit $A_1$ in the Step 2. Therefore, with suitable
$\Omega_{04A_2}(t)$ and $\Omega_{14A_2}(t)$, we can also make SQUID
qubit $A_2$ evolves from $|0\rangle_{A_2}$ to $|1\rangle_{A_2}$
($|1\rangle_{A_2}$ to $|0\rangle_{A_2}$) when SQUID qubit $B_2$ is
in state $|0\rangle_{B_2}$, while keep in $|0\rangle_{A_2}$ when
qubit $B_2$ is in state $|1\rangle_{B_2}$. After, Step 4 and Step 5,
the information for distinguishing $|\Psi_+\rangle_{B_1B_2}$
($|\Phi_+\rangle_{B_1B_2}$) from $|\Psi_-\rangle_{B_1B_2}$
($|\Phi_-\rangle_{B_1B_2}$) is encoded on SQUID qubit $A_2$.

\textbf{Step 6:} In this step, we only turn on classical microwave
fields $\Omega_{B_1}(t)$ and $\Omega_{B_2}(t)$ to perform inverse
transformations of Step 3 on SQUID qubits $B_1$ and $B_2$. When
\begin{eqnarray}\label{e15}
\int_{5T}^{6T}\Omega_{B_j}(t')dt'=-\pi/4,
\end{eqnarray}
the transformations for SQUID qubit $B_j$ are
\begin{eqnarray}\label{e16}
|0\rangle_{B_j}\rightarrow\frac{1}{\sqrt{2}}(|0\rangle_{B_j}-|1\rangle_{B_j}),\
\ \
|1\rangle_{B_j}\rightarrow\frac{1}{\sqrt{2}}(|0\rangle_{B_j}+|1\rangle_{B_j}).
\end{eqnarray}
After that, the states of SQUID qubits $B_1$ and $B_2$ recover to
their initial forms.

At the end of the Bell-state analysis, we detecting the states of
SQUID qubits $A_1$ and $A_2$, thus reading out the information for
distinguishing the four Bell states of SQUID qubits $B_1$ and $B_2$.
The measurement results of the states of SQUID qubits $A_1$ and
$A_2$ with corresponding Bell state of SQUID qubits $B_1$ and $B_2$
are shown in Table II. According to measurement results of the
states of SQUID qubits $A_1$ and $A_2$, a complete and
nondestructive Bell-state analysis can be realized.
\begin{center}
\centering{\bf Table II. Measurement results of the states of SQUID
qubits $A_1$ and $A_2$ with the corresponding Bell states of $B_1$
and $B_2$.} {\small\begin{tabular}{ccc} \hline\hline \ \ \
\ Measurement result\ \ \ \ &\ \ \ \ Corresponding Bell state\ \ \ \ \\
\hline
\ \ \ \ $|0\rangle_{A_1}|0\rangle_{A_2}$\ \ \ \ &\ \ \ \ $|\Psi_{+}\rangle_{B_1B_2}$\ \ \ \ \\
\ \ \ \ $|0\rangle_{A_1}|1\rangle_{A_2}$\ \ \ \ &\ \ \ \ $|\Psi_{-}\rangle_{B_1B_2}$\ \ \ \ \\
\ \ \ \ $|1\rangle_{A_1}|0\rangle_{A_2}$\ \ \ \ &\ \ \ \ $|\Phi_{+}\rangle_{B_1B_2}$\ \ \ \ \\
\ \ \ \ $|1\rangle_{A_1}|1\rangle_{A_2}$\ \ \ \ &\ \ \ \ $|\Phi_{-}\rangle_{B_1B_2}$\ \ \ \ \\
\hline \hline
\end{tabular}}
\end{center}

\section{Pulse design via STA}

In this section, let us design the microwave pulse via STA. As we
can see from Sec. III, the effective Hamiltonians of Step 1, Step 2,
Step 4 and Step 5 of the Bell-state analysis have the form
\begin{eqnarray}\label{e17}
H(t)=\Omega_1(t)|\psi_1\rangle\langle\psi_2|+\Omega_2(t)|\psi_2\rangle\langle\psi_3|+H.c..
\end{eqnarray}
The required transformation is
$|\psi_1\rangle\leftrightarrow|\psi_3\rangle$. Here, to build up
evolution paths, we consider the transitionless tracking algorithm.
As pointed out by Ref. \cite{CYHSR6} that, while utilizing
transitionless tracking algorithm, not only the eigenstates of the
Hamiltonian but also a set of orthonormalized time-dependent
vectors, can be selected as the evolution paths. For the current
protocol, we select
\begin{eqnarray}\label{e18}
&|\xi_1(t)\rangle&=(\cos\theta\sin\varphi\cos\vartheta+\sin\theta\sin\vartheta)|\psi_1\rangle+i\cos\varphi\cos\vartheta|\psi_2\rangle+(\sin\theta\sin\varphi\cos\vartheta-\cos\theta\sin\vartheta)|\psi_3\rangle,\cr\cr
&|\xi_2(t)\rangle&=(\cos\theta\sin\varphi\sin\vartheta-\sin\theta\cos\vartheta)|\psi_1\rangle+i\cos\varphi\sin\vartheta|\psi_2\rangle+(\sin\theta\sin\varphi\sin\vartheta+\cos\theta\cos\vartheta)|\psi_3\rangle,\cr\cr
&|\xi_3(t)\rangle&=\cos\theta\cos\varphi|\psi_1\rangle-i\sin\varphi|\psi_2\rangle+\sin\theta\cos\varphi|\psi_3\rangle,
\end{eqnarray}
where, $\theta$, $\varphi$ and $\vartheta$ are three time-dependent
parameters. According to transitionless tracking algorithm, the
evolution operator and the Hamiltonian for the evolution paths shown
in Eq.~(\ref{e18}) could be derived by
\begin{eqnarray}\label{e19}
&U'(t)&=\sum\limits_{n=1}^{3}|\xi_n(t)\rangle\langle\xi_n(0)|,
\end{eqnarray}
and
\begin{eqnarray}\label{e20}
&H'(t)&=i\sum\limits_{n=1}^{3}|\dot{\xi}_n(t)\rangle\langle\xi_n(t)|\cr\cr&&
=(\dot{\varphi}\cos\theta+\dot{\vartheta}\sin\theta\cos\varphi)|\psi_1\rangle\langle\psi_2|+(\dot{\varphi}\sin\theta-\dot{\vartheta}\cos\theta\cos\varphi)|\psi_2\rangle\langle\psi_3|\cr\cr&&
+i(\dot{\theta}-\dot{\vartheta}\sin\varphi)|\psi_3\rangle\langle\psi_1|+H.c..
\end{eqnarray}

To make $H'(t)=H(t)$, it requires
\begin{eqnarray}\label{e21}
&\Omega_1(t)&=\dot{\varphi}\cos\theta+\dot{\vartheta}\sin\theta\cos\varphi,\cr\cr
&\Omega_2(t)&=\dot{\varphi}\sin\theta-\dot{\vartheta}\cos\theta\cos\varphi,\cr\cr
&\dot{\theta}&=\dot{\vartheta}\sin\varphi.
\end{eqnarray}
Considering the time interval $[0,T]$ and the boundary condition
$\varphi(0)=-\pi/2$, $\varphi(T)=\pi/2$, $\vartheta(0)+\theta(0)=0$,
$\vartheta(T)-\theta(T)=\pi/2$, we have $U'(0)=I$ ($I$ is the
identical operator) and
$U'(T)=|\psi_1\rangle\langle\psi_3|+|\psi_3\rangle\langle\psi_1|+|\psi_2\rangle\langle\psi_2|$.
Thus, we can complete the transformation
$|\psi_1\rangle\leftrightarrow|\psi_3\rangle$. With the boundary
condition and Eq.~(\ref{e21}), parameters $\theta$, $\vartheta$ and
$\varphi$ can be selected to be
\begin{eqnarray}\label{e22}
\varphi=-\frac{\pi}{2}\cos(\frac{\pi t}{T}),\ \ \ \vartheta=\pi/4,\
\ \ \theta=-\pi/4.
\end{eqnarray}
Then, $\Omega_1(t)$ and $\Omega_2(t)$ could be derived as
\begin{eqnarray}\label{e23}
\Omega_1(t)=\frac{\pi^2}{2\sqrt{2}T}\sin(\pi t/T),\ \ \
\Omega_2(t)=-\frac{\pi^2}{2\sqrt{2}T}\sin(\pi t/T).
\end{eqnarray}

Therefore, if we set
\begin{eqnarray}\label{e24}
&&\Omega_{03A_1}(t)=\frac{\sqrt{g_{23A_1}^2+g_{B_1}^2}}{g_{B_1}}\Omega_1(t),\
\ \
\Omega_{13A_1}(t)=\frac{\sqrt{g_{23A_1}^2+g_{B_1}^2}}{g_{B_1}}\Omega_2(t),\cr\cr
&&\Omega_{04A_1}(t-T)=\frac{\sqrt{g_{24A_1}^2+g_{B_2}^2}}{g_{B_2}}\Omega_1(t),\
\
\Omega_{14A_1}(t-T)=\frac{\sqrt{g_{24A_1}^2+g_{B_2}^2}}{g_{B_2}}\Omega_2(t),\cr\cr
&&\Omega_{03A_2}(t-3T)=\frac{\sqrt{g_{23A_2}^2+g_{B_1}^2}}{g_{B_1}}\Omega_1(t),\
\
\Omega_{13A_2}(t-3T)=\frac{\sqrt{g_{23A_2}^2+g_{B_1}^2}}{g_{B_1}}\Omega_2(t),\cr\cr
&&\Omega_{04A_2}(t-4T)=\frac{\sqrt{g_{24A_2}^2+g_{B_2}^2}}{g_{B_2}}\Omega_1(t),\
\
\Omega_{14A_2}(t-4T)=\frac{\sqrt{g_{24A_2}^2+g_{B_2}^2}}{g_{B_2}}\Omega_2(t),
\end{eqnarray}
the microwave pulses could be used in the Bell-state analysis.

As for the $\Omega_{B_j}(t)$, which used to perform a single-qubit
operation on SQUID qubit $B_j$ in Step 3, according to
Eq.~(\ref{e12}), it could be chosen as
\begin{eqnarray}\label{e25}
\Omega_{B_j}(t-2T)=\frac{\pi^2}{8T}\sin(\pi t/T).
\end{eqnarray}
Similarly, for $\Omega_{B_j}(t)$, which used to perform a
single-qubit operation on SQUID qubit $B_j$ in Step 6, according to
Eq.~(\ref{e15}), it could be chosen as
\begin{eqnarray}\label{e26}
\Omega_{B_j}(t-5T)=-\frac{\pi^2}{8T}\sin(\pi t/T).
\end{eqnarray}

\begin{figure}
\scalebox{0.6}{\includegraphics[scale=1]{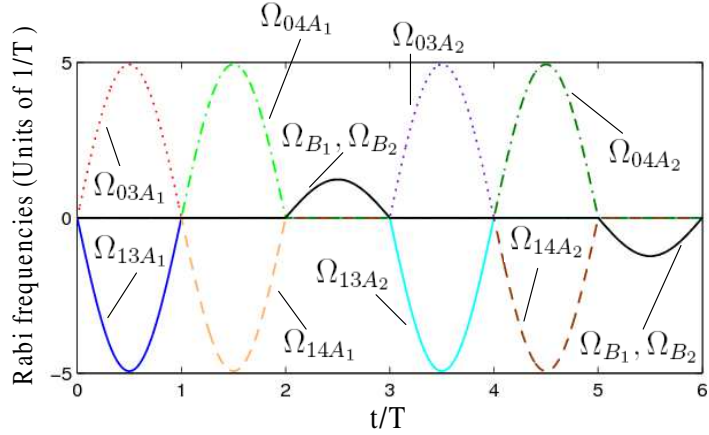}}
\caption{Microwave pulses for the Bell-state analysis.}\label{fig3}
\end{figure}

For simplicity, we consider $g_{23A_j}=g_{24A_j}=g_{B_j}=g$. The
microwave pulses for the Bell-state analysis is shown in
Fig.~\ref{fig3}. The maximal value of the amplitudes of the pulses
$\Omega_{\max}=\pi^2/2T\simeq4.9348/T$.

\section{Numerical simulations}

In this section, the robustness of the protocol is checked via
numerical simulations. Before the numerical simulations, we first
define the success probability for distinguishing each Bell state.
Assume that the density operator of the system is $\rho(t)$.
According to Table. II, the success probabilities could be defined
as
\begin{eqnarray}\label{e27}
&&P(\Psi_+)=\langle00\Psi_+|\rho(6T)|00\Psi_+\rangle,\ \
P(\Psi_-)=\langle01\Psi_-|\rho(6T)|01\Psi_-\rangle,\cr\cr
&&P(\Phi_+)=\langle10\Phi_+|\rho(6T)|10\Phi_+\rangle,\ \
P(\Phi_-)=\langle11\Phi_-|\rho(6T)|11\Phi_-\rangle,
\end{eqnarray}
where,
\begin{eqnarray}\label{e28}
&&|00\Psi_+\rangle=|0\rangle_{A_1}|0\rangle_{A_2}|\Psi_+\rangle_{B_1B_2},\
\
|01\Psi_-\rangle=|0\rangle_{A_1}|1\rangle_{A_2}|\Psi_-\rangle_{B_1B_2},\cr\cr
&&|10\Phi_+\rangle=|1\rangle_{A_1}|0\rangle_{A_2}|\Phi_+\rangle_{B_1B_2},\
\
|11\Phi_-\rangle=|1\rangle_{A_1}|1\rangle_{A_2}|\Phi_-\rangle_{B_1B_2}.
\end{eqnarray}

\begin{figure}
\scalebox{0.6}{\includegraphics[scale=1]{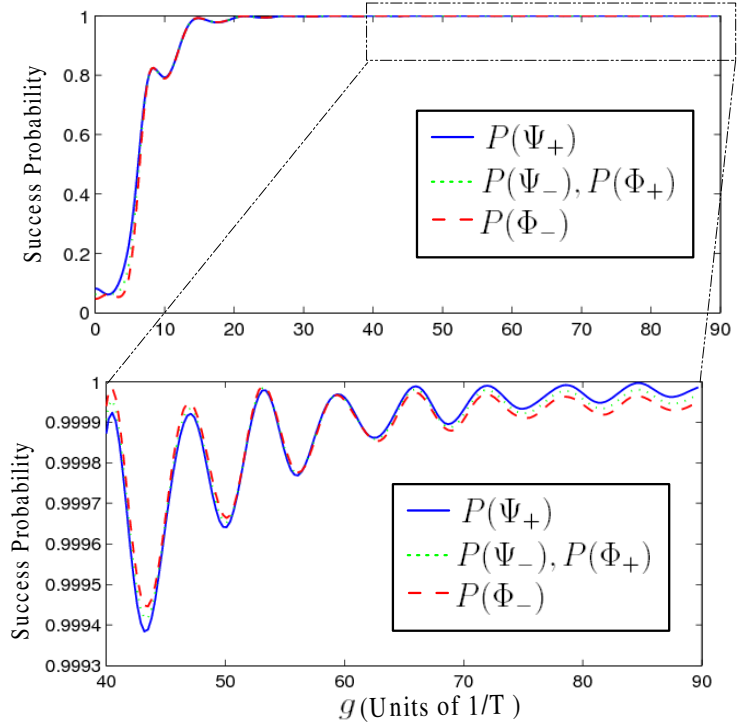}} \caption{Suceess
probabilities versus $g$. The solid-blue line: $P(\Psi_+)$; The
dotted-green line: $P(\Psi_-),P(\Phi_+)$; The dashed-red line
$P(\Phi_-)$.}\label{fig4}
\end{figure}

Firstly, as we perform the numerical simulation based on the
original Hamiltonian shown in Eq.~(\ref{e5}), a suitable coupling
constant $g$ should be chosen. Thus, we plot $P(\Psi_+)$,
$P(\Psi_-)$, $P(\Phi_+)$ and $P(\Phi_-)$ versus $g$ in
Fig.~\ref{fig4}. As shown in Fig.~\ref{fig4}, the success
probability are very low when $g$ is small due to bit flip errors,
while they are near 1 when $g\geq30/T$. According to
Fig.~\ref{fig4}, we choose $g=66/T$, which gives
$\max\{|1-P(\Psi_+)|,|1-P(\Psi_-)|,|1-P(\Phi_+)|,|1-P(\Phi_-)|\}\leq1.1847\times10^{-5}$.

\begin{figure}
\scalebox{0.6}{\includegraphics[scale=1]{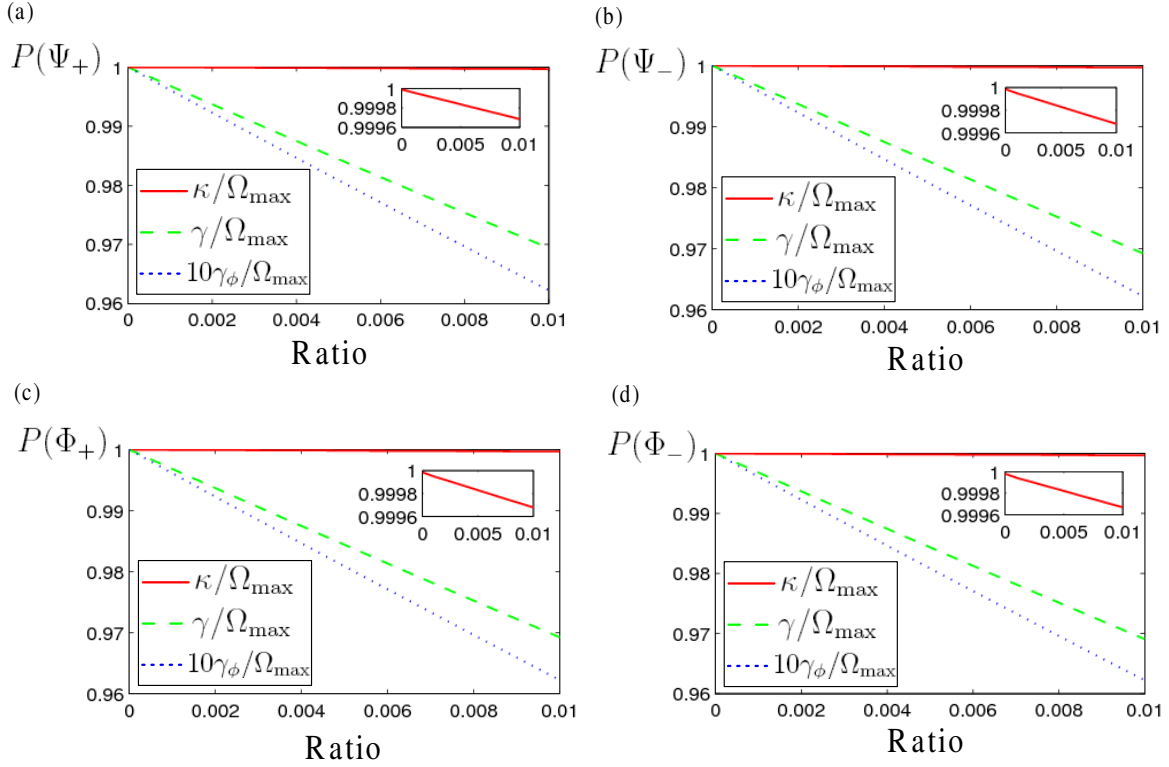}} \caption{(a)
$P(\Psi_+)$, (b) $P(\Psi_-)$, (c) $P(\Phi_+)$ and (d) $P(\Phi_-)$
versus $\kappa/\Omega_{\max}$ (the solid-red line),
$\gamma/\Omega_{\max}$ (the dashed-green line),
$10\gamma_\phi/\Omega_{\max}$ (the dotted-blue line).}\label{fig5}
\end{figure}

Secondly, we investigate the robustness of the protocol against
decoherence. The main decoherent factors for the protocol are: (i)
cavity dissipations for cavity fields $a_1$ and $a_2$ with decay
rates $\kappa_1$ and $\kappa_2$, respectively; (2) spontaneous
emissions from $|3\rangle_{A_j}$ ($|4\rangle_{A_j}$) to
$|0\rangle_{A_j}$, $|1\rangle_{A_j}$ and $|2\rangle_{A_j}$ with
spontaneous emission rates $\gamma_{30A_j}$ ($\gamma_{40A_j}$),
$\gamma_{31A_j}$ ($\gamma_{41A_j}$) and $\gamma_{32A_j}$
($\gamma_{42A_j}$), respectively; (3) spontaneous emissions from
$|2\rangle_{B_j}$ to $|0\rangle_{B_j}$ and $|1\rangle_{B_j}$ with
spontaneous emission rates $\gamma_{20B_j}$ and $\gamma_{21B_j}$,
respectively; (4) dephasings between $|3\rangle_{A_j}$
($|4\rangle_{A_j}$) and $|0\rangle_{A_j}$, $|1\rangle_{A_j}$,
$|2\rangle_{A_j}$ with dephasing rates $\gamma_{\phi30A_j}$
($\gamma_{\phi40A_j}$), $\gamma_{\phi31A_j}$ ($\gamma_{\phi41A_j}$),
$\gamma_{\phi32A_j}$ ($\gamma_{\phi42A_j}$), respectively; (5)
dephasings between $|2\rangle_{B_j}$ and $|0\rangle_{B_j}$,
$|1\rangle_{B_j}$ with spontaneous emission rates
$\gamma_{\phi20B_j}$, $\gamma_{\phi21B_j}$, respectively. Here, we
neglect the spontaneous emission and dephsing between
$|4\rangle_{A_j}$ and $|3\rangle_{A_j}$ and that between
$|2\rangle_{A_j}$ and $|0\rangle_{A_j}$ ($|1\rangle_{A_j}$), because
they are much weaker than spontaneous emissions and dephasings
between other levels of SQUID qubit $A_j$. Thus, the evolution of
the system could be described using a master equation
\begin{eqnarray}\label{e29}
\dot{\rho}(t)=i[\rho(t),H(t)]+\sum\limits_{p=1}^{34}[L_p\rho
L_p^{\dag}-\frac{1}{2}(L_p^{\dag}L_p\rho+\rho L_p^{\dag}L_p)],
\end{eqnarray}
where, $L_p$ ($p=1,2,3,...,34$) is the Lindblad operator as
\begin{eqnarray}\label{e30}
&&L_1=\sqrt{\gamma_{30A_1}}|0\rangle_{A_1}\langle3|,\ \
L_2=\sqrt{\gamma_{31A_1}}|1\rangle_{A_1}\langle3|,\ \
L_3=\sqrt{\gamma_{32A_1}}|2\rangle_{A_1}\langle3|,\cr\cr
&&L_4=\sqrt{\gamma_{40A_1}}|0\rangle_{A_1}\langle4|,\ \
L_5=\sqrt{\gamma_{41A_1}}|1\rangle_{A_1}\langle4|,\ \
L_6=\sqrt{\gamma_{42A_1}}|2\rangle_{A_1}\langle4|,\cr\cr
&&L_7=\sqrt{\gamma_{30A_2}}|0\rangle_{A_2}\langle3|,\ \
L_8=\sqrt{\gamma_{31A_2}}|1\rangle_{A_2}\langle3|,\ \
L_9=\sqrt{\gamma_{32A_2}}|2\rangle_{A_2}\langle3|,\cr\cr
&&L_{10}=\sqrt{\gamma_{40A_2}}|0\rangle_{A_2}\langle4|,\ \
L_{11}=\sqrt{\gamma_{41A_2}}|1\rangle_{A_2}\langle4|,\ \
L_{12}=\sqrt{\gamma_{42A_2}}|2\rangle_{A_2}\langle4|,\cr\cr
&&L_{13}=\sqrt{\gamma_{20B_1}}|0\rangle_{B_1}\langle2|,\ \
L_{14}=\sqrt{\gamma_{21B_1}}|1\rangle_{B_1}\langle2|,\ \
L_{15}=\sqrt{\gamma_{20B_2}}|0\rangle_{B_2}\langle2|,\cr\cr
&&L_{16}=\sqrt{\gamma_{21B_2}}|1\rangle_{B_2}\langle2|,\ \,
L_{17}=\sqrt{\kappa_1}a_1,\ \ L_{18}=\sqrt{\kappa_2}a_2\cr\cr
&&L_{19}=\sqrt{\gamma_{\phi30A_1}}(|3\rangle_{A_1}\langle3|-|0\rangle_{A_1}\langle0|),\
\
L_{20}=\sqrt{\gamma_{\phi31A_1}}(|3\rangle_{A_1}\langle3|-|1\rangle_{A_1}\langle1|),\cr\cr
&&L_{21}=\sqrt{\gamma_{\phi32A_1}}(|3\rangle_{A_1}\langle3|-|2\rangle_{A_1}\langle2|),\
\
L_{22}=\sqrt{\gamma_{\phi40A_1}}(|4\rangle_{A_1}\langle4|-|0\rangle_{A_1}\langle0|),\cr\cr
&&L_{23}=\sqrt{\gamma_{\phi41A_1}}(|4\rangle_{A_1}\langle4|-|1\rangle_{A_1}\langle1|),\
\
L_{24}=\sqrt{\gamma_{\phi42A_1}}(|4\rangle_{A_1}\langle4|-|2\rangle_{A_1}\langle2|),\cr\cr
&&L_{25}=\sqrt{\gamma_{\phi30A_2}}(|3\rangle_{A_2}\langle3|-|0\rangle_{A_2}\langle0|),\
\
L_{26}=\sqrt{\gamma_{\phi31A_2}}(|3\rangle_{A_2}\langle3|-|1\rangle_{A_2}\langle1|),\cr\cr
&&L_{27}=\sqrt{\gamma_{\phi32A_2}}(|3\rangle_{A_2}\langle3|-|2\rangle_{A_2}\langle2|),\
\
L_{28}=\sqrt{\gamma_{\phi40A_2}}(|4\rangle_{A_2}\langle4|-|0\rangle_{A_2}\langle0|),\cr\cr
&&L_{29}=\sqrt{\gamma_{\phi41A_2}}(|4\rangle_{A_2}\langle4|-|1\rangle_{A_2}\langle1|),\
\
L_{30}=\sqrt{\gamma_{\phi42A_2}}(|4\rangle_{A_2}\langle4|-|2\rangle_{A_2}\langle2|),\cr\cr
&&L_{31}=\sqrt{\gamma_{\phi20B_1}}(|2\rangle_{B_1}\langle2|-|0\rangle_{B_1}\langle0|),\
\
L_{32}=\sqrt{\gamma_{\phi21B_1}}(|2\rangle_{B_1}\langle2|-|1\rangle_{B_1}\langle1|),\cr\cr
&&L_{33}=\sqrt{\gamma_{\phi20B_2}}(|2\rangle_{B_2}\langle2|-|0\rangle_{B_2}\langle0|),\
\
L_{34}=\sqrt{\gamma_{\phi21B_2}}(|2\rangle_{B_2}\langle2|-|1\rangle_{B_2}\langle1|).
\end{eqnarray}
For brief discussions, we assume
$\gamma_{30A_j}=\gamma_{31A_j}=\gamma_{32A_j}=\gamma_{40A_j}=\gamma_{41A_j}=\gamma_{42A_j}=\gamma_{20B_j}=\gamma_{21B_j}=\gamma$,
$\gamma_{\phi30A_j}=\gamma_{\phi31A_j}=\gamma_{\phi32A_j}=\gamma_{\phi40A_j}=\gamma_{\phi41A_j}=\gamma_{\phi42A_j}=\gamma_{\phi20B_j}=\gamma_{\phi21B_j}=\gamma_\phi$,
$\kappa_1=\kappa_2=\kappa$. $P(\Psi_+)$, $P(\Psi_-)$, $P(\Phi_+)$
and $P(\Phi_-)$ versus $\gamma/\Omega_{\max}$,
$10\gamma_\phi/\Omega_{\max}$ and $\kappa/\Omega_{\max}$ are plotted
in Figs.~\ref{fig5}~(a-d), respectively.
As shown by Fig.~\ref{fig5}, the Bell-state analysis is quite robust
against the cavity decays. The success probabilities of
distinguishing each Bell state are still higher than 0.9996 when
$\kappa/\Omega_{\max}=0.01$. We can also find that, the Bell-state
analysis is more sensitive to the spontaneous emissions of SQUID
qubits. When $\gamma/\Omega_{\max}=0.01$, the success probabilities
of distinguishing each Bell state are all a little less than 0.97.
As for the dephasings, they are the most troublesome decoherent
factors. When $\gamma_\phi/\Omega_{\max}=0.001$, the success
probabilities of distinguishing each Bell state are all higher than
0.96. We considered experimentally realizable parameters
$\Omega_{\max}=2\pi\times6.37$MHz, $g=2\pi\times85.14$MHz,
$\kappa=1.32$MHz, $\gamma=0.40$MHz, $\gamma_\phi=0.20$MHz
\cite{YCPPRA67,YCPPRL92,XZLRMP85}, we have $P(\Psi_+)=0.9555$,
$P(\Psi_-)=0.9554$, $P(\Phi_+)=0.9554$ and $P(\Phi_-)=0.9554$.
Therefore, the protocol possesses high success probability to
distinguish each one of four Bell states.

\section{Conclusion}

In conclusion, we have proposed a protocol for complete Bell-state
analysis for superconducting-quantum-interference-device qubits.
Although the Bell-state analysis is composed of six steps, one just
need to use a sequence of sinusoidal microwave pulses as shown in
Fig.~(\ref{fig3}), without any additional operations. Therefore, the
operations are not difficult in experiments. After the six steps,
the information for distinguishing four Bell states are encoding on
two auxiliary qubits $A_1$ and $A_2$. Thus, we can read out the
information by detecting the states of $A_1$ and $A_2$. The
detections of SQUID qubits have been reported in protocols
\cite{BennettSST20,TakayanagiSM32}. In this paper, we have not
considered the detection efficiency, while for a real experiment,
they should be taken into account.  Moreover, apart from the
advantages of SQUID qubits, the protocol holds some other
advantages:

(1) The Bell-state analysis is complete and nondestructive. In other
words, we can completely distinguish four Bell states without
destroying them. Thus the physical resource could be saved.

(2) As shown by the numerical simulations, under the current
experimental conditions, the protocol still possess high success
probability of distinguishing each Bell state when decoherence is
taken into account.

(3) Since STA is used in the pulse design, the operation speed of
the protocol may be faster than that using the adiabatic passages.

Thus, we hope the protocol could contribute to information readout
for quantum communications and quantum computations in
superconducting quantum networks.

\section*{Acknowledgement}

This work was supported by the National Natural Science Foundation
of China under Grants No. 11575045, No. 11374054 and No. 11674060,
and the Major State Basic Research Development Program of China
under Grant No. 2012CB921601.

\section*{Appendix: The derivation of the effective Hamiltonian}

Here, we would like to amply described the derivation of the
effective Hamiltonian $H_{eff1}(t)$ shown in Eq.~(\ref{e9}) from
Eqs. (\ref{e7}) and (\ref{e8}). According to the eigenstates of
$H_{c1}$ shown in Eq.~(\ref{e8}), we can rewrite Eq.~(\ref{e7}) as
\begin{eqnarray}\label{e31}
&H_{step1}(t)&=H_{m1}(t)+H_{c1},\cr\cr
&H_{m1}(t)&=\frac{\Omega_{03A_1}(t)}{\sqrt{2(g_{23A_1}^2+g_{B_1}^2})}|\bar{\psi}_1\rangle(\sqrt{2}g_{B1}\langle\bar{\phi}_0|+g_{23A_1}\langle\bar{\phi}_3|+g_{23A_1}\langle\bar{\phi}_4|)\cr\cr&&
+\frac{\Omega_{03A_1}(t)}{\sqrt{2}}|\bar{\psi}_6\rangle(\langle\bar{\phi}_1|+\langle\bar{\phi}_2|)+\frac{\Omega_{13A_1}(t)}{\sqrt{2}}|\bar{\psi}_9\rangle(\langle\bar{\phi}_1|+\langle\bar{\phi}_2|)\cr\cr&&
+\frac{\Omega_{13A_1}(t)}{\sqrt{2(g_{23A_1}^2+g_{B_1}^2})}|\bar{\psi}_5\rangle(\sqrt{2}g_{B1}\langle\bar{\phi}_0|+g_{23A_1}\langle\bar{\phi}_3|+g_{23A_1}\langle\bar{\phi}_4|)+H.c.,\cr\cr
&H_{c1}&=g_{23A_1}(|\bar{\phi}_1\rangle\langle\bar{\phi}_1|-|\bar{\phi}_2\rangle\langle\bar{\phi}_2|)+\sqrt{g_{23A_1}^2+g_{B_1}^2}(|\bar{\phi}_3\rangle\langle\bar{\phi}_3|-|\bar{\phi}_4\rangle\langle\bar{\phi}_4|).
\end{eqnarray}
Considering $H_{c1}$ as a free Hamiltonian, we perform the picture
transformation $\bar{U}=e^{-iH_{c1}t}$ on Eq.~(\ref{e31}). The
Hamiltonian after the transformation is
\begin{eqnarray}\label{e32}
&H'_{step1}(t)&=\frac{\Omega_{03A_1}(t)}{\sqrt{2(g_{23A_1}^2+g_{B_1}^2})}|\bar{\psi}_1\rangle(\sqrt{2}g_{B1}\langle\bar{\phi}_0|+g_{23A_1}\langle\bar{\phi}_3|e^{-i\sqrt{g_{23A_1}^2+g_{B_1}^2}t}+g_{23A_1}\langle\bar{\phi}_4|e^{i\sqrt{g_{23A_1}^2+g_{B_1}^2}t})\cr\cr&&
+\frac{\Omega_{13A_1}(t)}{\sqrt{2(g_{23A_1}^2+g_{B_1}^2})}|\bar{\psi}_5\rangle(\sqrt{2}g_{B1}\langle\bar{\phi}_0|+g_{23A_1}\langle\bar{\phi}_3|e^{-i\sqrt{g_{23A_1}^2+g_{B_1}^2}t}+g_{23A_1}\langle\bar{\phi}_4|e^{i\sqrt{g_{23A_1}^2+g_{B_1}^2}t})\cr\cr&&
+\frac{\Omega_{03A_1}(t)}{\sqrt{2}}|\bar{\psi}_6\rangle(\langle\bar{\phi}_1|e^{-ig_{23A_1}t}+\langle\bar{\phi}_2|e^{ig_{23A_1}t})+\frac{\Omega_{13A_1}(t)}{\sqrt{2}}|\bar{\psi}_9\rangle(\langle\bar{\phi}_1|e^{-ig_{23A_1}t}+\langle\bar{\phi}_2|e^{ig_{23A_1}t})\cr\cr&&+H.c..
\end{eqnarray}
Under the condition $\Omega_{03A_1}(t),\Omega_{13A_1}(t)\ll
g_{23A_1},g_{B_1}$, we neglect the terms of high frequency
oscillations and obtain the effective Hamiltonian
\begin{eqnarray}\label{e33}
H_{eff1}(t)=\frac{g_{B_1}}{\sqrt{g_{23A_1}^2+g_{B_1}^2}}[\Omega_{03A_1}(t)|\bar{\psi}_1\rangle\langle\bar{\phi}_0|+\Omega_{13A_1}(t)|\bar{\psi}_5\rangle\langle\bar{\phi}_0|]+H.c..
\end{eqnarray}

In similar way, we can derive the effective Hamiltonian
$H_{eff2}(t)$ shown in Eq.~(\ref{e11}) from Eqs. (\ref{e9}) and
(\ref{e10}).

\end{document}